\documentclass[12pt]{article}
\usepackage{graphicx}
\begin{document}

\title{Time evolution of tunneling and decoherence: \\
soluble model.}

\author{A. N. Salgueiro$^{(1)}$, A.F.R. de Toledo Piza$^{(1)}$ \\ 
J. G. Peixoto de Faria$^{(2)}$ and M.C. Nemes$^{(2)}$}

\maketitle

\begin{center}

{$^{(1)}$ Departamento de F\'{\i}sica--Matem\'atica,
 Instituto de F\'\i sica,
 Universidade de S\~ao Paulo, \\ C.P. 66318, 
 CEP 05315-970 S\~ao Paulo, S.P., Brazil}

{$^{(2)}$ Departamento de F\'\i sica, ICEX, 
 Universidade Federal de Minas Gerais,\\ C.P. 702, CEP 
 30161-970 Belo Horizonte, M.G. Brazil}

\end{center}

\begin{abstract}

Decoherence effects associated to the damping of a tunneling two-level
system are shown to dominate the tunneling probability at short times
in strong coupling regimes in the context of a soluble model. A
general decomposition of tunneling rates in dissipative and unitary
parts is implemented. Master equation treatments fail to describe the
model system correctly when more than a single relaxation time is
involved.

\end{abstract}


\section{Introduction}

Motion in tunneling phenomena can be seen quite generally as resulting
from dispersive effects in the time evolution of spatially localized
quantum states which are not energy eigenstates. Since, from this
point of view, quantum interference plays a central role in these
phenomena, one may ask how are they affected by quantum decoherence
processes which occur as a result of coupling to other degrees of
freedom. Effects of additional degrees of freedom in tunneling
processes have been studied now for a long time and in a variety of
contexts \cite{gen}, but emphasis falls usually on the question of the
observable changes on inclusive properties such as tunneling
probabilities. However, these questions are often addressed in
semi-classical terms, including generalizations of the WKB approach
\cite{BNW}, which tend to becloud the dispersive character of the
quantum dynamics.

In order to analyze the dispersive dynamics of tunneling we consider in
this paper the simple but not unrealistic case of two-level
approximations to symmetric, bound, bi-stable systems whose energy
spectrum is characterized by a structure of doublets. Each member of
one doublet consists of symmetric/antisymmetric superpositions of
localized states which may be seen as coupled by effective tunneling
amplitudes. In the two-level approximation this is conveniently
expressed in terms of a spin-1/2 algebra. Dissipation and decoherence
phenomena can be introduced in this description by coupling the
two-level system to another system, possibly with a continuous energy
spectrum. A collection of harmonic oscillators is frequently used in
this connection, leading to the ``spin-boson'' model analyzed by
Leggett and collaborators \cite{legg}. There the effects of
dissipation on the tunneling rates are studied independently of
explicit control of properties of the tunneling state relating to
decoherence processes. A simpler, soluble model, which allows for a
control of this sort, results when the two-level system is suitably
coupled to a generic additional system with continuous spectrum, as
described in detail in section \ref{main} below. In this model the
coupled dynamics can be reduced to the problem of the spreading of a
product ``doorway state'' due to its coupling to a continuous
background also of product states, which allows for a treatment in
terms of the techniques developed long ago by Fano \cite{fano} in the
context of atomic scattering theory. While this schematic model is
less ``realistic'' from the point of view of energy-loss mechanisms,
it retains the overall dynamical features of the spin-boson model
relating to coherence-loss, which is our main concern here. The model
allows for the calculation of the reduced density associated with the
two-level system at all times. Model dependences are indicated by
differences in the results obtained in the two cases. We find, in
particular, that transition rates out of a localized initial state can
be substantially enhanced by decoherence in strong coupling regimes.

The use of Master equations for the description of damped two-level
systems is well known \cite{livopt}. The usual derivation of these
equations \cite{meq} assume weak-coupling to an harmonic oscillator
reservoir and high temperature, so that their use in strong coupling
and zero temperature situations is not warranted. Comparison of the
exact model results with solutions of these Master equations agree for
particular forms of the coupling independently of the overall coupling
strength and at zero temperature. In general, however, the
perturbative ingredients of the Master equation derivations appear as
limiting factors.

\section{Dissipative and unitary tunneling rates.}\label{Pt}

The quantum state of a two-level subsystem can be generally described
in terms of a (time dependent) hermitian reduced density operator of
unit trace $\rho(t)$. The simple two-level effective Hamiltonian

\[
H_\sigma=\frac{\epsilon}{2}\,(\sigma_3+\hat{1}_\sigma),
\]

\noindent where $\hat{1}_\sigma$ stands for the unit $2\times 2$
matrix, can be understood as describing the tunneling between the
``localized'' (``left'' and ``right'') states

\[
|l\rangle\equiv\frac{|+\rangle+|-\rangle}{\sqrt{2}}\hspace{2cm}
{\rm and}\hspace{2cm}|r\rangle\equiv\frac{|+\rangle-|-\rangle}
{\sqrt{2}}
\]

\noindent where $|\pm\rangle$ are the eigenvectors of the diagonal
Pauli matrix $\sigma_3$, and hence also energy eigenstates. The time
dependent density matrix corresponding to the initial condition

\begin{equation}
\label{incond}
\rho(t=0)=|l\rangle\langle l|
\end{equation}

\noindent is given in the $|\pm\rangle$ basis as (with $\hbar=1$)

\begin{equation}
\label{pure}
\rho(t)=\frac{1}{2}\left(\begin{array}{cc} 1 & e^{i\epsilon t} \\
e^{-i\epsilon t} & 1 \end{array}\right)=\rho^2(t).
\end{equation}
 
\noindent As is well known, the property of idempotency is equivalent
to the purity of the quantum state described by $\rho(t)$. A tunneling
probability $P(t)$ can be defined as the probability of finding this
system in the complementary localized state $|r\rangle$ at time $t$,
and is readily calculated as

\begin{equation}
\label{ptun}
P(t)={\rm Tr}\left[|r\rangle\langle
r|\rho(t)\right]=\sin^2\frac{\epsilon t}{2}.
\end{equation}

When the two-level subsystem is coupled to additional degrees of
freedom its state at time $t$ is still described by a reduced density
matrix which is non-negative, hermitian and of unit trace, but in
general {\it not} idempotent. In fact, the quantity $\delta(t)=1-{\rm
Tr}[\rho(t)^2]$ (referred to as the {\it linear entropy} or the {\it
idempotency defect}) is often used as a measure of the decoherence
suffered by the two level subsystem through its coupling to the other
degrees of freedom.

A most convenient way of representing such a reduced density matrix
makes use of its eigenvalues and eigenvectors

\[
\rho(t)|t,k\rangle=p_k(t)|t,k\rangle,\hspace{.5cm}k=1,2;\hspace{1cm}
\langle t,k|t,k'\rangle=\delta_{kk'},\hspace{.5cm}0\leq p_k(t)\leq 1
\]

\noindent in terms of which one can write

\begin{equation}
\label{natorb}
\rho(t)=\sum_k |t,k\rangle p_k(t)\langle t,k|.
\end{equation}

\noindent It should be noted that in general both the eigenvectors and
the eigenvalues in (\ref{natorb}) are time-dependent. While the
eigenvectors evolve unitarily (as they define a time-dependent
orthonormal basis in the quantum phase-space of the two-level
subsystem) the time-dependence of the eigenvalues $p_k(t)$ reveal the
change in time of the coherence properties of the state of the
two-level subsystem. The idempotency defect is in fact given in terms
of the eigenvalues $p_k(t)$ as

\begin{equation}
\label{delta}
\delta=1-\sum_k p_k(t)^2=2p_1(t)(1-p_1(t))
\end{equation}

\noindent where use has been made of the unit trace property of
$\rho(t)$. In the case of eq. (\ref{pure}), it can be easily checked
that $p_1=1$ and $p_2=0$ at all times, so that $\delta\equiv
0$. Assuming that the reduced density is that which evolves from the
initial condition (\ref{incond}), the tunneling probability is now
given by

\[
P(t)=\sum_k p_k(t)|\langle r|t,k\rangle|^2.
\]

\noindent In view of what has been said concerning the time-dependence
of the ingredients of this expression, it is immediately clear that
the tunneling {\it rate}

\[
R(t)\equiv\dot{P}(t)=\sum_k\dot{p}_k(t)|\langle r|t,k\rangle|^2+
\sum_k p_k(t)\frac{d}{dt}|\langle r|t,k\rangle|^2
\]

\noindent splits into two contributions which can be unambiguously
ascribed to changes in the coherence properties of the state of the
subsystem and to its unitary evolution, respectively:

\begin{equation}
\label{Rd}
R_d(t)\equiv\sum_k\dot{p}_k(t)|\langle r|t,k\rangle|^2=\left(2
|\langle r|t,1\rangle|^2-1\right)\dot{p}_1(t)
\end{equation}

\noindent and

\begin{equation}
\label{Ru}
R_u(t)\equiv\sum_k p_k(t)\frac{d}{dt}|\langle r|t,k\rangle|^2=
\left(2p_1(t)-1\right)\frac{d}{dt}|\langle r|t,1\rangle|^2.
\end{equation}

\noindent The last forms of $R_u$ and $R_d$ make use of the unit trace
of $\rho$ and also of the normalization of $|r\rangle$. In the case of
the purely unitary time evolution under the simple two-level
Hamiltonian $H_\sigma$ one has of course $R_d(t)\equiv 0$ and
$R_u(t)=R(t)=(\epsilon/2)\sin\,\epsilon t$. This decomposition of
transition rates in dissipative and unitary parts can be simply
extended to cases involving reduced densities of larger
dimensionality, as indicated in Appendix A. 

\section{Two-level tunneling with coupling to a
continuum.}\label{main} 

A simple model allowing for decoherence effects in the two-level
tunneling processes is one in which the two-level subsystem described
by the Hamiltonian $H_\sigma$ is coupled to an additional
``nondescript'' set of degrees of freedom (which will be identified by
means of a subscript $b$) to which is associated an energy spectrum
containing one discrete state represented by a normalized state vector
$|0_b\rangle$ in addition to a continuum of states $|\eta\rangle$,
normalized in energy as $\langle\eta|\eta'\rangle=\delta(\eta-\eta')$.
The model is characterized further by the Hamiltonian

\begin{eqnarray}
\label{Hmod}
H&=&\frac{\epsilon}{2}(\sigma_3+\hat{1}_\sigma)\otimes{\bf{\hat{1}}}_b
+\left(|0_b\rangle e_0\langle 0_b|+\int_{\eta_0}^{\bar{\eta}}d\eta\,
|\eta\rangle\eta\langle\eta|\right)\otimes\hat{1}_\sigma \nonumber \\
&&+\frac{1}{2}\int_{\eta_0}^{\bar{\eta}}d\eta\,\left[g(\eta)|\eta
\rangle\sigma_-\langle 0_b|+g^*(\eta)|0_b\rangle\sigma_+\langle\eta|
\right]\\ &&+\frac{1}{2}\int_{\eta_0}^{\bar{\eta}}d\eta\,\left[g'(
\eta)|\eta\rangle\sigma_+\langle 0_b|+g'^*(\eta)|0_b\rangle\sigma_-
\langle\eta|\right]\nonumber
\end{eqnarray}

\noindent where $g(\eta)$ and $g'(\eta)$ are coupling matrix elements
$\langle -\eta|\hat{g}|+0_b\rangle$ and $\langle+\eta|\hat{g}'|-0_b
\rangle$ respectively, possibly dependent on $\eta$. In the special
case $g(\eta)=g'(\eta)$ the coupling terms reduce to

\[
\sigma_x\int_{\eta_0}^{\bar{\eta}}d\eta\;\left[|\eta\rangle g(\eta)
\langle 0_b|+|0_b\rangle g^*(\eta)\langle\eta|\right]
\]

\noindent which resembles in its structure the coupling term of the
spin-boson model\cite{legg}.

A basis in the quantum phase-space of the composite system described
by equation (\ref{Hmod}) can be constructed as $\{|s0_b\rangle,\,\{|
s\eta\rangle\}\},\;\,s=+,-,\;\,\eta_0\leq\eta \leq\bar{\eta}$, and the
adopted structure of the coupling terms shows that the subspaces
spanned by the two sets of basis vectors $\{|+0_b\rangle,\,\{|-\eta
\rangle\}\}$ and $\{|-0_b\rangle,\,\{|+ \eta\rangle\}\}$ are closed
under the action of $H$. Different choices for the model parameters
$\epsilon$, $e_0$, the span of the continuum band $\eta_0$ to
$\bar{\eta}$ and the $\eta$-dependence of the coupling functions $g$
and $g'$ allow for a considerable variety of dynamical
regimes. Closest to the situation described by the spin-boson model is
that in which $e_0=\eta_0=0$ with $g(\eta)=g^*(\eta)=g'(\eta)$. Some
aspects of the relationship of these two models are briefly discussed
in Appendix B.  The particular version of the model which will be used
in the following sections differs from this, however. A manifest
feature of this version is the assumption that one has

\[
\eta_0\ll e_0\hspace{1cm}{\rm and}\hspace{1cm}
e_0+\epsilon\ll\bar{\eta}
\]

\noindent so that the two product states $|\pm0_b\rangle$, with
eigen-energies $e_0$ and $e_0+\epsilon$ fall well within a
sufficiently wide continuum band of product states $|\pm\eta\rangle$.
This feature will in fact allow for the continuum spreading of the
states $|+0_b\rangle$ and $|-0_b\rangle$ through their coupling with
states $|-\eta\rangle$ and $|+\eta\rangle$ respectively with spreading
widths appreciably larger than the doublet splitting $\epsilon$, a
situation which will be seen to characterize strong damping
regimes. Furthermore, since these two spreadings are dynamically
independent, as they take place in different invariant subspaces of
$H$, there will be in general two distinct relaxation times
associated with the damped two-level subsystem.  

General state vectors in each of the invariant subspaces of $H$ can be
written as

\begin{eqnarray*}
|\psi_a\rangle&=&a_0|+0_b\rangle+\int_{\eta_0}^{\bar{\eta}}d\eta\,
A(\eta)|-\eta\rangle \\
|\psi_b\rangle&=&b_0|-0_b\rangle+\int_{\eta_0}^{\bar{\eta}}d\eta\,
B(\eta)|+\eta\rangle.
\end{eqnarray*}

\noindent In terms of these expansions, the eigenvalue problem for $H$

\[
(E-H)|\psi_{a,b}^{(E)}\rangle=0
\]

\noindent can be reduced to two independent pairs of coupled equations
with similar structure for the expansion coefficients

\begin{eqnarray}
\label{forw}
(E-\epsilon-e_0)a_0^{(E)}&=&\int_{\eta_0}^{\bar{\eta}}d\eta\,
g^*(\eta)A^{(E)}(\eta), \nonumber \\
(E-\eta)A^{(E)}(\eta)&=&g(\eta)a_0^{(E)}
\end{eqnarray}

\noindent and

\begin{eqnarray}
\label{backw}
(E-e_0)b_0^{(E)}&=&\int_{\eta_0}^{\bar{\eta}}d\eta\,
g'^*(\eta)B^{(E)}(\eta), \nonumber \\
(E-\epsilon-\eta)B^{(E)}(\eta)&=&g'(\eta)b_0^{(E)}.
\end{eqnarray}

\noindent The solutions of these coupled equations has been given long
ago by Fano\cite{fano} and van Kampen\cite{vankampen}. In the case of
eqs. (\ref{forw}) one has

\begin{equation}
\label{a0ls}
|a_0^{(E)}|^2=\frac{|g(E)|^2}{[E-\epsilon-e_0-F(E)]^2+\pi^2|g(E)|^4}
\end{equation}

\noindent with $F(E)$ given by the principal value integral

\begin{equation}
\label{shift}
F(E)={\cal{P}}\int_{\eta_0}^{\bar{\eta}}d\eta\,\frac{|g(\eta)|^2}
{E-\eta}.
\end{equation}

\noindent The continuum coefficients are

\begin{equation}
\label{Afroma0}
A^{(E)}(\eta)=\left[\frac{{\cal{P}}}{E-\eta}+z(E)\delta(E-\eta)\right]
g(\eta)a_0^{(E)}
\end{equation}

\noindent with

\[
z(E)=\frac{E-\epsilon-e_0-F(E)}{|g(E)|^2}.
\]

\noindent The value of the amplitude $a_0^{(E)}$ involves a phasing
convention which is subsequently carried over to the $A^{(E)}(\eta)$
through equation (\ref{Afroma0}). The solution of eqs. (\ref{backw})
is of course completely analogous and does not have to be given
explicitly here.

Strictly speaking, when the integration limits of the continuum
variable $\eta$ are {\it finite}, the spectrum of $H$ may include
discrete states outside the continuum range, subject to the
$\eta$-dependence of the coupling parameters $g(\eta)$. Whenever
needed, the discrete eigenvalues of $H$ can be obtained (e.g. in the
case of eqs. (\ref{forw})) as solutions $E_d<\eta_0$ or
$E_d>\bar{\eta}$ of the dispersion equation

\[
E-\epsilon-e_0=\int_{\eta_0}^{\bar{\eta}}d\eta\,\frac{|g(\eta)|^2}
{E-\eta}.
\]

\noindent In this case eq. (\ref{Afroma0}) is replaced by

\[
A^{(E_d)}(\eta)=\frac{g(\eta)}{E_d-\eta}
\]

\noindent and $a_0^{(E)}$ is determined from the normalization
condition (up to an arbitrary phase factor) as

\[
|a_0^{(E_d)}|^2=\left(1-\left(\frac{dF(E)}{dE}\right)_{E=E_d}
\right)^{-1}.
\]

\noindent The low-energy discrete solutions are specially relevant for
comparison of this model with other models of the damping mechanism,
such as the many-oscillators model (see Appendix B).

\subsection{Time evolution of a localized initial
condition.}\label{tev} 

Using the stationary states of $H$ as determined above one can write
the time evolution of the localized initial state

\begin{equation}
\label{tunic}
|t=0\rangle=\frac{|+\rangle+|-\rangle}{\sqrt{2}}\otimes|0_b\rangle=
|l\rangle\otimes|0_b\rangle.
\end{equation}

\noindent Note that in the absence of coupling, $g(\eta)=g'(\eta)=0$,
the time evolution reduces to that found in section \ref{Pt}. In
general, the component states $|+0_b\rangle$ and $|-0_b\rangle$ can be
written in terms of the stationary states determined from the solutions
of eqs. (\ref{forw}) and (\ref{backw}), so that

\begin{eqnarray*}
|t\rangle=e^{-iHt}|t=0\rangle&=&\frac{e^{-iHt}}{\sqrt{2}}\left(
\sum_{E_d}a_0^{(E_d)}|\psi_a^{(E_d)}\rangle+\int_{\eta_0}^{\bar{\eta}
}dE\;a_0^{(E)*}|\psi_a^{(E)}\rangle \right. \\ &+& \left.
\sum_{E'_d}b_0^{(E'_d)}|\psi_b^{(E'_d)}\rangle+\int_{\eta_0}^{\bar
{\eta}}dE\;b_0^{(E)*}|\psi_a^{(E)}\rangle\right)
\end{eqnarray*}

\noindent When the continuum range extends over a sufficiently broad
energy interval on both sides of $e_0$ the discrete state amplitudes
$a_0^{(E_d)}$ and $ b_0^{(E'_d)}$ become very small and can be
neglected. In what follows we assume this to be the case in order to
avoid unessential complications. By re-expressing the stationary
states in terms of the factorized bases, forming the total density
$|t\rangle\langle t|$ and taking its trace over the $b$-states one
obtains for the reduced density of the two level system at time $t$

\[
\rho(t)=\left(\begin{array}{cc} \rho_{++}(t) & \rho_{+-}(t) \\ &
\\ \rho_{+-}^*(t) & 1-\rho_{++}(t) \end{array}\right)
\]

\noindent with

\begin{equation}
\label{rho++}
\rho_{++}(t)=\frac{1}{2}\left(1+\left|\int_{\eta_0}^{\bar{\eta}}dE\,
|a_0^{(E)}|^2e^{-iEt}\right|^2-\left|\int_{\eta_0}^{\bar{\eta}}dE\,
|b_0^{(E)}|^2e^{-iEt}\right|^2\right)
\end{equation}

\noindent and

\begin{eqnarray}
\label{rho+-}
\rho_{+-}(t)&=&\frac{1}{2}\left(\int_{\eta_0}^{\bar{\eta}}dE\,
|a_0^{(E)}|^2e^{-iEt}\int_{\eta_0}^{\bar{\eta}}dE'\,|b_0^{(E')}|^2
e^{-iE't} \right. \nonumber \\ &&+ \left.
\int_{\eta_0}^{\bar{\eta}}dE\int_{\eta_0}^{\bar{\eta}}dE'
b_0^{(E)*}a_0^{(E')}e^{-i(E-E')t}\int_{\eta_0}^{\bar{\eta}}d\eta\,
B^{(E)}(\eta)A^{(E')*}(\eta)\right).
\end{eqnarray}

\noindent The last term of eq. (\ref{rho+-}) is the only one in which
the continuum amplitudes $A^{(E')}(\eta)$ and $B^{(E)}(\eta)$
intervene explicitly. This term describes a re-correlation of
components of the state of the two-level system in the {\it single}
continuum subspace of system $b$, to which they couple through
$g(\eta)$ and $g'(\eta)$. In fact, a simple variant of the model that
avoids this re-correlation process can be devised by adding a second
continuum $\{|\zeta\rangle\}$ orthogonal to the first, i.e. $\langle
\eta|\zeta\rangle\equiv 0$, to which the component $|-0_b\rangle$ is
coupled through $g'(\zeta)$. It is easy to see that the resulting
expressions for $\rho_{++}$ and $\rho_{+-}$ are in this case identical
to eqs. (\ref{rho++}) and (\ref{rho+-}) except for the absence of the
last term in the latter.

The tunneling probability at time $t$ of the two-level subsystem, from
the initial state $|l\rangle$ to the complementarily localized
state $|r\rangle$, defined in terms of the reduced density matrix
$\rho(t)$ as in eq. (\ref{ptun}), is generally given as

\begin{equation}
\label{ptgen}
P(t)={\rm Tr}\left[|r\rangle\langle r|\rho(t)\right]=\frac{1}{2}
\left(1-2\,{\rm Re}\,\rho_{+-}\right)
\end{equation}

\noindent and the idempotency defect (\ref{delta}) of the reduced
density is

\begin{equation}
\label{delgen}
\delta(t)=2\left[\rho_{++}(t)\left(1-\rho_{++}(t)\right)-
|\rho_{+-}(t)|^2\right]=2\det\rho(t).
\end{equation}

\subsection{Special case: single relaxation time.}\label{onel}

An interesting particular case is that in which $g'(\eta)\equiv 0$, so
that $|-0_b\rangle$ is a stationary state of $H$ while $|+0_b\rangle$
is spread through its coupling to the $|-\eta\rangle$ continuum. In
this case the reduced density matrix elements are simply given in
terms of the spectral distribution $|a_0^{(E)}|^2$, eq. (\ref{a0ls}),
of the state $|+0_b\rangle$ as

\begin{equation}
\label{rho++gp0}
\rho_{++}(t)\rightarrow\frac{1}{2}\left|\int_{\eta_0}^{\bar{\eta}}dE\,
|a_0^{(E)}|^2e^{-iEt}\right|^2,\hspace{1cm}g'(\eta)\equiv 0
\end{equation}

\noindent and

\begin{equation}
\label{rho+-gp0}
\rho_{+-}(t)\rightarrow\frac{1}{2}\int_{\eta_0}^{\bar{\eta}}dE\,
|a_0^{(E)}|^2e^{-i(E-e_0)t},\hspace{1cm}g'(\eta)\equiv 0.
\end{equation}

\noindent These expressions can be immediately evaluated in closed
form when the matrix elements $g(\eta)$ are independent of $\eta$, and
the energy shift $F(E)$, eq. (\ref{shift}) is slowly energy dependent
over an interval of the order of $\pi|g|^2$ in the neighborhood of
$E=\epsilon+e_0$. This will in fact be the case when the continuum
range extends over sufficiently broad intervals both above and below
this energy. Equation (\ref{a0ls}) is then well approximated by the
Breit-Wigner form

\[
|a_0^{(E)}|^2\rightarrow\frac{|g|^2}{(E-e_R)^2+\pi^2|g|^4}\equiv
\frac{1}{2\pi}\;\frac{\Gamma}{(E-e_R)^2-\frac{\Gamma^2}{4}}
\]

\noindent where $e_R-\epsilon-e_0-F(e_R)=0$ and $\Gamma=2\pi|g|^2$ is
the usual width parameter of the Breit-Wigner distribution, given in
Golden Rule form. Note that for large $\bar{\eta}-e_0\simeq
e_0-\eta_0$ one has $F(e_R)\simeq F(e_0)\simeq 0$ so that
$e_R-e_0\simeq\epsilon$. Extending the integration limits to
$\pm\infty$ one obtains by simple contour integrations

\begin{equation}
\label{ols}
\rho_{++}(t)=\frac{1}{2}e^{-\Gamma t}\hspace{1.5cm}{\rm and}
\hspace{1.5cm}\rho_{+-}(t)=\frac{1}{2}e^{-i(e_R-e_0)t-
\frac{\Gamma t}{2}}
\end{equation}

\noindent so that

\begin{equation}
\label{pt+}
P(t)=\frac{1}{2}\left(1-e^{-\frac{\Gamma t}{2}}
\cos(e_R-e_0)t\right).
\end{equation}

\noindent It should be kept in mind that the model assumptions allow
$\Gamma > \epsilon$, in which case the initial time-dependence of
$P(t)$ is dominated by the decaying exponential factor. The
decoherence process undergone by the reduced density as measured by
the idempotency defect (\ref{delta}) is given by

\begin{equation}
\label{del+}
\delta(t)=\frac{1}{2}e^{-\Gamma t}\left(1-e^{-\Gamma t}\right)
\end{equation}

\noindent so that the tunneling and the decoherence {\it rates}

\[
\dot{P}(t)=\frac{e^{-\frac{\Gamma t}{2}}}{2}\left(\frac{\Gamma}{2}
\cos(e_R-e_0)t+(e_R-e_0)\sin(e_R-e_0)t\right)
\]

\noindent and

\[
\dot{\delta}(t)=\frac{\Gamma}{2}e^{-\Gamma t}\left(2e^{-\Gamma t}-1
\right)
\]

\noindent are both proportional to the damping width $\Gamma$, related
to the single relaxation time $\tau_d$ defined as $\tau_d=1/\Gamma$.
Moreover, $\dot{P}(t)$ contains the period of unitary evolution
$\tau_u\equiv 2\pi/(e_R-e_0)\simeq 2\pi/\epsilon$ as an additional,
independent time scale. In the strong coupling regime
$\tau_d\ll\tau_u$ eq. (\ref{pt+}) shows dominance of coherence loss
effects in $P(t)$ for times $t<\tau_d$ leading eventually to
saturation at $P=1/2$. The sorting of dissipative and unitary
contributions to the tunneling rate $\dot{P}(t)$ is achieved through
the calculation of the rates $R_d(t)$ and $R_u(t)$ introduced in
section \ref{Pt}. As shown in eqs. (\ref{Rd}) and (\ref{Ru}), this
involves the time dependence of the eigenvalues and eigenvectors of
the reduced density. The relevant ingredients can be expressed in a
straightforward way in terms of the reduced density matrix elements
$\rho_{++}(t)$ and $\rho_{+-}(t)$.  Since the resulting expressions
are lengthy and not particularly illuminating they will not be given
here. Numerical results illustrating these features, obtained using
the matrix elements given in eq. (\ref{ols}), are shown in figs. 1 and
2 respectively for the weak and strong coupling regimes. The figures
show $P(t)$, $\delta(t)$ and the tunneling rates $\dot{P}(t)$,
$R_d(t)$ and $R_u(t)$. The correlation of $\delta(t)$ with $R_d(t)$
and the dominance of the latter over $R_u(t)$ in the strong coupling
regime are clearly seen.

\subsection{General case: two relaxation times.}\label{twol}

When both $g(\eta)$ and $g'(\eta)$ are different from zero the matrix
elements of the reduced densities are given by the complete
expressions (\ref{rho++}) and (\ref{rho+-}). Using the tunneling
initial condition of subsection \ref{tev} and under the assumptions
used in subsection \ref{onel} the first of these can be evaluated to
yield

\begin{equation}
\label{2lrho++}
\rho_{++}(t)=\frac{1}{2}\left(1+e^{-\Gamma t}-e^{-\Gamma' t}\right)
\end{equation}

\noindent where $\Gamma'$, defined as the width parameter $\Gamma'
=2\pi|g'|^2$ ($g'$ having been assumed to be independent of $\eta$) of
the distribution

\[
|b_0^{(E)}|^2=\frac{1}{2\pi}\;\frac{\Gamma'}{(E-e'_R)^2+
\frac{\Gamma'^2}{4}},
\]

\noindent introduces an additional time scale in the evolution of the
composite system. As for $\rho_{+-}$ one gets, from eq. (\ref{rho+-}),
using eq. (\ref{Afroma0}) and the methods of ref. \cite{fano}

\begin{equation}
\label{2lrho+-}
\rho_{+-}(t)=\frac{1}{2}\left[e^{-i(e_R-e'_R)t-\frac{\Gamma+\Gamma'}
{2}t}+\frac{2\sqrt{\Gamma\Gamma'}}{\Gamma+\Gamma'}e^{i(e_R-e'_R)t}
\left(1-e^{-\frac{\Gamma+\Gamma'}{2}t}\right)\right]
\end{equation}

\noindent where the last term is due to the re-correlation processes
mentioned in subsection \ref{tev}. Note that under conditions in which
the shift functions $F(e_R)$ and $F'(e'_R)$ are small one has
$e_R-e'_R\simeq\epsilon$. This will be assumed to be the case from
here on in order to save notation.

The tunneling probability is given by eq. (\ref{ptgen}) as

\begin{equation}
\label{pt2gamma}
P(t)=\frac{1}{2}-\frac{\cos\epsilon t}{2}\left[\frac{2\sqrt{\Gamma
\Gamma'}}{\Gamma+\Gamma'}+e^{-\frac{\Gamma+\Gamma'}{2}t}\left(1-
\frac{2\sqrt{\Gamma\Gamma'}}{\Gamma+\Gamma'}\right)\right]
\end{equation}

\noindent which shows that the re-correlation processes give rise to
steady state oscillations of $P(t)$ after the transient associated
with the decaying exponential has died down. The amplitude of the
steady-state oscillations depends on the damping widths $\Gamma$ and
$\Gamma'$, however, and is maximum when $\Gamma=\Gamma'$, in which
case the transient actually cancels out and the undamped form of
$P(t)$, eq. (\ref{ptun}), is recovered.   

This last result does not imply, however, that there is no decoherence
when $\Gamma=\Gamma'$. The general expression for $\delta(t)$ obtained
from eq. (\ref{delgen}) can be written as

\begin{eqnarray*}
\delta(t)&=&\frac{1}{2}\left(1+e^{-\Gamma t}-e^{-\Gamma' t}\right)
\left(1-e^{-\Gamma t}+e^{-\Gamma' t}\right) - \frac{1}{2}e^{-(\Gamma+
\Gamma')t}\\ &&-\frac{2\sqrt{\Gamma\Gamma'}}{\Gamma+\Gamma'}\left(1-
e^{-\frac{\Gamma+\Gamma'}{2}t}\right)\left[e^{-\frac{\Gamma+\Gamma'}
{2}t}\cos2\epsilon t+\frac{\sqrt{\Gamma\Gamma'}}{
\Gamma+\Gamma'}\left(1-e^{-\frac{\Gamma+\Gamma'}{2}t}\right)\right]
\end{eqnarray*}

\noindent and reduces, when $\Gamma=\Gamma'$, to

\[
\delta(t)\hspace{.5cm}\stackrel{\Gamma=\Gamma'}{\longrightarrow}
\hspace{.5cm} 2e^{-\Gamma t}\left(1-e^{-\Gamma t}\right)
\sin^2\epsilon t
\]

\noindent which vanishes only asymptotically, as $t\rightarrow
\infty$. In the general case $\Gamma\neq\Gamma'$ with $\Gamma,
\Gamma'\neq 0$, one has the asymptotic limit

\[
\lim_{t\rightarrow\infty}\delta(t)=\frac{1}{2}-\frac{2\Gamma\Gamma'
}{(\Gamma+\Gamma')^2}.
\]

\noindent The restriction $\Gamma,\Gamma'\neq 0$ is necessary since
the limit $t\rightarrow\infty$ does not commute with the limit
$\Gamma'\rightarrow 0$. If the alternate version of the model with two
orthogonal continua is used, all terms involving factors
$\sqrt{\Gamma\Gamma'}/{(\Gamma+ \Gamma')}$ are absent, both in $P(t)$
and in $\delta(t)$. In this case the two-level system decoheres
maximally when $t\rightarrow\infty$ as two orthogonal components of
the two-level subsystem become correlated to orthogonal continuum
wave packets.

The dissipative and unitary tunneling rates $R_d$ and $R_u$ can also be
obtained from the reduced density in a straightforward way. A typical
result for moderately strong coupling is shown in fig. 3.

\subsection{Interpretation of model results.}

In spite of the schematic character of the model, the features
emerging from its analysis underline relevant aspects which should be
kept in mind in the study of more realistic cases. As shown explicitly
e.g. in the general equation (\ref{ptgen}), the coherence properties
directly relevant for localization are, in the representation used
there, contained in the {\it real part} of the matrix element
$\rho_{+-}$. The harmonic time dependence of this quantity in the
absence of any external entanglements is modified when these effects
are turned on by making $g$ and $g'$ (or, equivalently, $\Gamma$ and
$\Gamma'$) different from zero, as can be read from the time dependent
term of equation (\ref{pt2gamma}). In the case considered in section
\ref{onel} ($\Gamma\neq 0$, $\Gamma'=0$) the harmonic oscillations are
damped by the effects of the external coupling, leading, in the strong
coupling limit, to tunneling rates dominated by the associated
relaxation times, rather that by the oscillation period. When also
$\Gamma'\neq 0$, on the other hand, a new undamped oscillatory term
arises due to contributions to ${\rm Re}\,\rho_{+-}$ coming from
continuum components, an effect which has been referred to as a
re-correlation process in section \ref{twol}. The amplitude of the
damped component is at the same time reduced, leading to its complete
cancellation in the limiting case $\Gamma=\Gamma'\neq 0$.

It is important to stress that the coherence properties bearing on the
purity of the complete tunneling {\it state}, which can be measured in
a basis-independent way in terms of the determinant of the reduced
density matrix (see equation (\ref{delgen})), involve other dynamic
quantities besides the localization-specific correlation part ${\rm
Re}\,\rho_{+-}$, and therefore in general correlate poorly with the
tunneling probability $P(t)$. In particular, when $\Gamma=\Gamma'\neq
0$ the transient decoherence undergone by the tunneling state has no
effect on the tunneling probability, since the localization-specific
coherence properties remain the same as in the absence of the external
couplings due to the re-correlation process. The decomposition of the
tunneling rate $\dot{P}(t)$ into dissipative and unitary parts
$R_d(t)$ and $R_u(t)$ can thus be seen as expressing the dependence
of the time rate of change of the localization-specific correlation
part ${\rm Re}\,\rho_{+-}$ in terms of contributions associated to the
(unitary) time evolution of the eigenvectors and from the
(non-unitary) time evolution of the eigenvalues of the reduced density
which describes the tunneling state. Only the latter contributions
relate to the overall decoherence of the tunneling state.

\section{Master equation results.}

The dissipative dynamics of the reduced density matrix of a two-level
system is often described in terms of a master equation valid in a
weak-coupling regime, derived \cite{meq} under the assumption of
linear coupling to a reservoir of harmonic oscillators, which is
subsequently eliminated from the description under the assumption of
the validity of a Born-Markov approximation. In addition, one assumes
in the derivation that correlation functions involving reservoir
degrees of freedom can be meaningfully evaluated for all times using
the initial state of the reservoir, usually taken to be a thermal
equilibrium state. The usual form of the master equation \cite{livopt}
is obtained assuming that the coupling to the reservoir is given by

\[
H_{\rm int}=\sum_k \left(g_k b_k^\dagger\sigma_-+g_k^* b_k\sigma_+
\right)
\]

\noindent where $b_k$, $b_k^\dagger$ are the lowering and rising
operators for the $k$-th bath oscillator. This coupling term is
similar to the term involving $g(\eta)$ in eq. (\ref{Hmod}) if one
associates the ground state of the reservoir with the discrete state
$|0_b\rangle$ of the model. The resulting form of the equation is
then, in the limit of zero temperature,

\[
\dot{\rho}(t)=-i\frac{\epsilon}{2}\left[\sigma_3,\rho(t)\right]+
\frac{\gamma}{2}\left(2\sigma_-\rho(t)\sigma_+-\sigma_+\sigma_-
\rho(t)-\rho(t)\sigma_+\sigma_-\right)
\]

\noindent where $\gamma$ is a damping coefficient related to $H_{\rm
int}$ and to the distribution of reservoir frequencies. 

The solution of this equation satisfying the initial condition

\begin{equation}
\label{icm}
\rho_{++}(0)=\rho_{+-}(0)=\frac{1}{2}
\end{equation}

\noindent is

\[
\rho_{++}(t)=\frac{1}{2}e^{-\gamma t}\hspace{1cm}{\rm and}
\hspace{1cm}\rho_{+-}(t)=\frac{1}{2}e^{i\epsilon t-\frac{\gamma}{2} t}
\]

\noindent which reproduces the result (\ref{ols}), with the
correspondence $\gamma\leftrightarrow\Gamma$, up to details related to
the shift function $F(E)$ in that case. It should be kept in mind,
however, that the derivation of the master equation is limited to weak
coupling regimes due to the Born-Markov approximation.

One may next try and include effects analogous to those produced by
the coupling term involving $g'(\eta)$ in the model Hamiltonian
(\ref{Hmod}) by adding a new term to $H_{\rm int}$ in the derivation
of the master equation. Thus, using now

\[
H_{\rm int}\rightarrow\sum_k \left(g_k b_k^\dagger\sigma_-+g_k^* b_k
\sigma_+\right)+\sum_k \left(g'_k b_k^\dagger\sigma_+ +{g'}_k^* b_k
\sigma_-\right)
\]

\noindent we obtain, following the standard derivation procedure
\cite{meq}, the modified master equation

\begin{eqnarray}
\label{cme}
\dot{\rho}(t)&=&-i\frac{\epsilon}{2}\left[\sigma_3,\rho(t)\right]+
\frac{\gamma}{2}\left(2\sigma_-\rho(t)\sigma_+-\sigma_+\sigma_-
\rho(t)-\rho(t)\sigma_+\sigma_-\right) \\ &&+\frac{\gamma'}{2}\left(2
\sigma_+\rho(t)\sigma_--\sigma_-\sigma_+\rho(t)
-\rho(t)\sigma_-\sigma_+\right)+\frac{\sqrt{\gamma\gamma'}}{2}
\left(2\sigma_+\rho(t)\sigma_++2\sigma_-\rho(t)\sigma_-\right)
\nonumber
\end{eqnarray}

\noindent where $\gamma'$ a the constant analogous to $\gamma$ but
related to the new terms added to $H_{\rm int}$. The solution of this
equation satisfying the same initial condition (\ref{icm}) is

\[
\rho_{++}(t)=\frac{1}{2}e^{-(\gamma+\gamma')t}+\frac{\gamma'}
{\gamma+\gamma'}\left(1-e^{-(\gamma+\gamma')t}\right)
\]

\noindent and

\[
\rho_{+-}(t)=\frac{e^{-(\gamma+\gamma')t}}{2\sqrt{\epsilon^2-\gamma
\gamma'}}\left[\sqrt{\gamma\gamma'}-i\epsilon\,{\rm sen}\left(2
\sqrt{\epsilon^2-\gamma\gamma'}\;t\right)+\sqrt{\epsilon^2-\gamma
\gamma'}\cos\left(2\sqrt{\epsilon^2-\gamma\gamma'}\;t\right)\right]
\]

\noindent which is different from the corresponding results obtained
in subsection \ref{twol}. In particular, the model result for
$\rho_{++}(t)$, eq. (\ref{2lrho++}), involves two relaxation times
associated to the damping widths $\Gamma$ and $\Gamma'$ appearing in
the difference of separate exponentials, while the master equation
result involves just one ``effective'' relaxation time appearing in a
single exponential involving the sum $\gamma+\gamma'$. The relation
between these two results can be revealed by expanding both solutions
to {\it first order} in the damping parameters. One obtains, from the
master equation result,

\[
\rho_{++}(t)\rightarrow\frac{1}{2}-\frac{\gamma}{2}t+
\frac{\gamma'}{2}t
\]

\noindent which is identical to what one obtains from
eq. (\ref{2lrho++}) with the replacements $\gamma\leftrightarrow
\Gamma$ and $\gamma'\leftrightarrow\Gamma'$. The expansion of the
master equation result for $\rho_{+-}(t)$ coincides likewise with the
expansion of eq. (\ref{2lrho+-}). These facts can be understood by
recalling that a Born-Markov approximation is involved in the
derivation of the master equation. The Born approximation implies a
truncation of an expansion in the couplings, producing a result which
is subsequently ``exponentiated'' to all orders through the Markov
assumption. As a result of this there is no general guarantee for
results of the master equation which go beyond the order of the
truncation involved in the Born approximation. In particular, the
dynamics leading to two different relaxation times is not properly
accounted for. Note, in this connection, that when $\gamma=\gamma'$
(i.e. when the two relaxation times coincide) the master equation
result reproduces the model result up to a small shift of the bare
frequency $\epsilon$, if one restricts oneself to weak coupling
regimes.

It may be argued that the master equation (\ref{cme}) is derived on
the assumption of a reservoir of harmonic oscillators, which differs
from the dynamical assumptions of the model defined by the Hamiltonian
(\ref{Hmod}). However, the same master equation also follows from this
Hamiltonian, with the usual (in fact, even somewhat weaker) derivation
assumptions, establishing, in particular, the correspondence
$\gamma\leftrightarrow\Gamma$ and $\gamma'\leftrightarrow\Gamma'$.
The Born approximation has to be performed in the same way in both
cases, indicating again that it is at the root of the discussed
discrepancies.

\section{Concluding remarks.}

The main general conclusion to be drawn from the analysis of the model
described by the Hamiltonian (\ref{Hmod}) with localized initial
condition (\ref{tunic}) is that a clear distinction should be made
between coherence properties relating a) to localization in the
tunneling degree of freedom and b) to the degree of quantum purity of
the reduced density (\ref{natorb}) describing the quantum state of the
tunneling system, as measured e.g. by the linear entropy
(\ref{delta}). Coherence properties of class a) are carried by the
real part of off-diagonal matrix element $\rho_{+-}(t)$ in the basis
which diagonalizes $H_\sigma$ or, equivalently, by the diagonal
matrix elements in the localized basis $\{|r\rangle,|l\rangle\}$,
e.g. $\rho_{rr}(t)=1/2-{\rm Re}\, \rho_{+-}=P(t)$. Coherence
properties of class b), on the other hand, are carried by the
eigenvalues of the reduced density, and involve therefore other
dynamic quantities in addition to that which is relevant for class
a).

The contribution of decoherence processes undergone by the reduced
density to the coherence properties of class a) can be isolated
through the general decomposition of the tunneling rate $\dot{P}(t)$
into dissipative and unitary parts, eqs. (\ref{Ru}) and (\ref{Rd}).
Results for the damped two-level model show that the dissipative
component contributes most at times short on the scale of the bare
frequency $\epsilon$. The enhanced initial tunneling rate in the
strong coupling case with no re-correlation effects (see
fig. \ref{fig2}) is dominated by the dissipative component, and can
thus be interpreted in terms of a large effect of the coherence loss
of the tunneling state also on the localization-specific correlation
properties of the initial state of the two-level subsystem.

Comparison of the exact solution of the model with master equation
results derived in the usual way from the model Hamiltonian indicates
that master equations appear to be inadequate to deal with situations
involving more than a single relaxation time. This difficulty can be
ascribed to the Born approximation involved in the derivation of these
equations.

\vspace{.3cm}

\noindent {\bf Acknowledgements.} JGPF and MCN are partially supported
by Conselho Nacional de Desenvolimento Cient\'{\i}fico e Tecnol\'ogico
(CNPq), Brazil. ANS is supported by Funda\c{c}\~ao de Amparo à
Pesquisa do Estado de S\~ao Paulo (FAPESP). MCN has been partly
supported by FAPESP during the later stages of this work.

\section*{Appendix A.}

A general reduced density matrix can always be written in terms of its
time dependent natural orbitals $|n(t)\rangle$ and associated
eigenvalues $p_n(t)$ as

\[
\rho(t)=\sum_n|n(t)\rangle p_n(t)\langle n(t)|.
\]

\noindent The probability that the state thus described is found in a
given (time independent) subspace defined in terms of a projector

\[
{\cal{P}}\equiv\sum_k |b_k\rangle\langle b_k|,
\]

\noindent where the $|b_k\rangle$ constitute an orthonormal set of
state vectors, is given by

\[
P(t)={\rm Tr}\left[\rho(t){\cal{P}}\right].
\]

\noindent The vectors $|b_k\rangle$ can be expanded in the natural
orbitals of $\rho(t)$ as $|b_k\rangle=\sum_n\beta_n^{(k)}(t)|n(t)
\rangle$ so that $P(t)$ is expressed as

\[
P(t)=\sum_n p_n(t)\;\sum_k|\beta_n^{(k)}|^2.
\]

\noindent Derivation with respect to time then gives dissipative and
unitary contributions proportional to $\dot{p}_n$ and
$\dot{\beta}_n^{(k)}$ respectively.

\section*{Appendix B.}

In view of the widespread use of coordinate coupling to a collection
of harmonic oscillators in order to introduce damping effects on the
dynamics of simple quantum-mechanical systems (including two-level
systems), and since the results obtained in this way seem to be at
variance with those of the continuum model for the damping as used in
section \ref{main}, this Appendix deals briefly with a comparison of
features of the two models. Instead of the implementation which
minimizes ``threshold'' effects adopted in that section, the case
$e_0\simeq\eta_0\simeq 0$ with $g(\eta)=g^*(\eta)=g'(\eta)$ of the
continuum model will be considered, together with with the ``zero
bias'' spin-boson model \cite{legg}.

In both cases the Hamiltonian can be split as $H_\sigma+H_b+H_{int}$
with $H_\sigma=\frac{\epsilon}{2} \sigma_3$ and

\[
H_b=|0_b\rangle e_0\langle 0_b|+\int_{\eta_0}^{\bar{\eta}}d\eta\,
|\eta\rangle\eta\langle\eta|,\hspace{1cm}H_{int}=\sigma_1\int_{
\eta_0}^{\bar{\eta}}d\eta\,g(\eta)\left(|\eta\rangle
\langle0_b|+|0_b\rangle\langle\eta|\right)\equiv\sigma_1 G
\]

\noindent for the continuum model of equation (\ref{Hmod}) and

\[
H_{b}=\sum_\alpha \hbar\omega_\alpha a_\alpha^\dagger a_\alpha,
\hspace{1cm}H_{int}=\sigma_1\sum_\alpha g_\alpha\sqrt{\frac{\hbar}
{2m_\alpha\omega_\alpha}}\left(a_\alpha^\dagger+a_\alpha\right)
\equiv\sigma_1 G.
\]

\noindent for the spin-boson model. A useful way of representing these
operators uses the spin basis of localized states in which $\sigma_1$
is diagonal. The Hamiltonian is then represented in both cases as

\[
H=\left(\begin{array}{cc}H_b+G & -\frac{i}{2}\epsilon \\ & \\
\frac{i}{2}\epsilon & H_b-G \end{array}\right).
\]

\noindent The operators $H_b\pm G$ can be separately diagonalized
exactly in the two cases. This solves the problem completely in the
extreme adiabatic limit $\epsilon\rightarrow 0$. In general, matrix
elements in the off-diagonal blocks involving $H_\sigma$ are modified
by overlap factors involving eigenstates of the two operators $H_b\pm
G$. Rather than dealing with the full problem, we restrict ourselves
to the perturbative effects of $H_b$ for the lowest adiabatic
states. These are, in both cases, a degenerate doublet which we denote
by $|\phi_0^{\pm}\rangle$, with energy $E_0$. The secular matrix to be
considered is then

\[
\left(\begin{array}{cc} E_0 & -\frac{i}{2}\epsilon\langle\phi_0^{(+)}|
\phi_0^{(-)}\rangle \\ & \\ \frac{i}{2}\epsilon\langle\phi_0^{(-)}|
\phi_0^{(+)}\rangle & E_0 \end{array}\right).
\]

In the case of the spin-boson model the overlap $\langle\phi_0^{(+)}|
\phi_0^{(-)}\rangle$ involves oscillator coherent states with opposite
displacements which depend, in particular, on the coupling constants
$g_\alpha$. The ensuing reduction of the overlap implies the reduction
of the splitting of the perturbed states, favoring the slowing down of
transition rates out of localized states. In the case of the continuum
model, on the other hand, $E_0$ is the root which lies below $\eta_0$
of the dispersion equation

\[
E_0-e_0=\int_{\eta_0}^{\bar{\eta}}d\eta\;\frac{g^2(\eta)}{E_0-\eta}.
\]

\noindent The corresponding unperturbed states can be expanded as

\[
|\phi_0^{(\pm)}\rangle=a_0^{(\pm)}|0_b\rangle+\int_{\eta_0}^{\bar{
\eta}}d\eta\;A^{(\pm)}(\eta)|\eta\rangle
\]

\noindent the expansion coefficients being given by

\[
|a_0^{(\pm)}|^2=\frac{1}{1+\frac{1}{4}\int_{\eta_0}^{\bar{\eta}}
d\eta\;\frac{g^2(\eta)}{(E_0-\eta)^2}}\hspace{1cm}{\rm and}
\hspace{1cm}A^{(\pm)}(\eta)=\pm\frac{g(\eta)}{E_0-\eta}\,a_0^{(\pm)}.
\]

\noindent and the overlap factor is

\[
\langle\phi_0^{(+)}|\phi_0^{(-)}\rangle=\frac{1-\frac{1}{4}\int_{
\eta_0}^{\bar{\eta}}d\eta\;\frac{g^2(\eta)}{(E_0-\eta)^2}}{1+
\frac{1}{4}\int_{\eta_0}^{\bar{\eta}}d\eta\;\frac{g^2(\eta)}{
(E_0-\eta)^2}}.
\]

\noindent The behavior of this object involves the energy dependence
of the coupling function $g(\eta)$ and of the solution $E_0$ of the
dispersion equation. In the case when $g^2(\eta)=\gamma\eta^s$, with
$0\leq s<1$ one finds, for fixed $e_0=\eta_0=0$ that $E_0$ approaches
minus infinity and $|a_0^{(\pm)}|^2$ approaches unity as the high
energy cut-off $\bar{\eta}$ increases. In the same limit the overlap
also approaches unity and the transition rate out of the localized
states approaches its free value. For values of $\bar{\eta}$ not too
large in comparison with $\epsilon$ the overlap is smaller than one,
leading to a quenching of this transition rate. These behaviors are
supported by exact numerical results.  


\newpage

\section*{Figure captions.}

\noindent Fig. 1 - Tunneling probability $P(t)$, idempotency defect
$\delta(t)$ (top) and tunneling rates $R_d(t)$, $R_u(t)$ and
$\dot{P}(t)=R_d(t)+R_u(t)$ (bottom) in the single finite relaxation
time, weak-coupling case with parameters $\Gamma=0.3$ and
$\epsilon=1$. The time scale is defined so that $\hbar=1$.

\vspace{.5cm}

\noindent Fig. 2 - Same as Figure 1 in the single finite relaxation
time, strong coupling case with parameters $\Gamma=3.$, $\epsilon=1$.
Note the smaller depth of the time scale in this case.

\vspace{.5cm} 

\noindent Fig. 3 - Same as Figure 1 in the case of two finite
relaxation times with parameters $\Gamma=2$, $\Gamma'=.5$,
$\epsilon=1$.

\newpage

\begin{figure}
\centering
\begin{minipage}[c]{.45\textwidth}
\centering
\caption{Tunneling probability $P(t)$, idempotency defect $\delta(t)$
(top) and tunneling rates $R_d(t)$, $R_u(t)$ and $\dot{P}(t)=R_d(t)+
R_u(t)$ (bottom) in the single finite relaxation time, weak-coupling
case with parameters $\Gamma=0.3$ and $\epsilon=1$. The time scale is
defined so that $\hbar=1$.}
\label{fig1}
\end{minipage}%
\hfill
\begin{minipage}[c]{.45\textwidth}
\begin{center}
\includegraphics[width=2.7in]{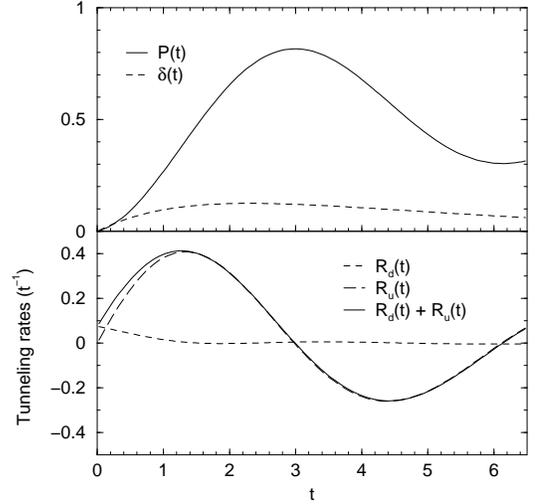}
\end{center}
\end{minipage}
\end{figure}

\begin{figure}
\centering
\begin{minipage}[c]{.45\textwidth}
\centering
\caption{Same as Figure \ref{fig1} in the of single finite relaxation
time, strong coupling case with parameters $\Gamma=3.$, $\epsilon=1$.
Note the smaller depth of the time scale in this case.}
\label{fig2}
\end{minipage}%
\hfill
\begin{minipage}[c]{.45\textwidth}
\begin{center}
\includegraphics[width=2.7in]{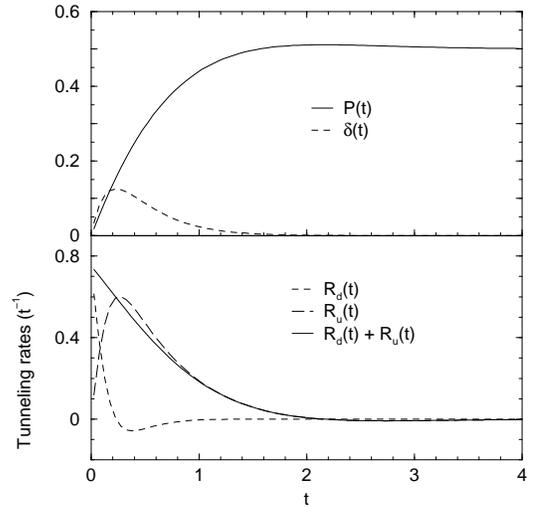}
\end{center}
\end{minipage}
\end{figure}

\begin{figure}
\centering
\begin{minipage}[c]{.45\textwidth}
\centering
\caption{Same as Figure \ref{fig1} in the case of two finite
relaxation times, with parameters $\Gamma=2$, $\Gamma'=.5$,
$\epsilon=1$.}
\label{fig3}
\end{minipage}%
\hfill
\begin{minipage}[c]{.45\textwidth}
\begin{center}
\includegraphics[width=2.7in]{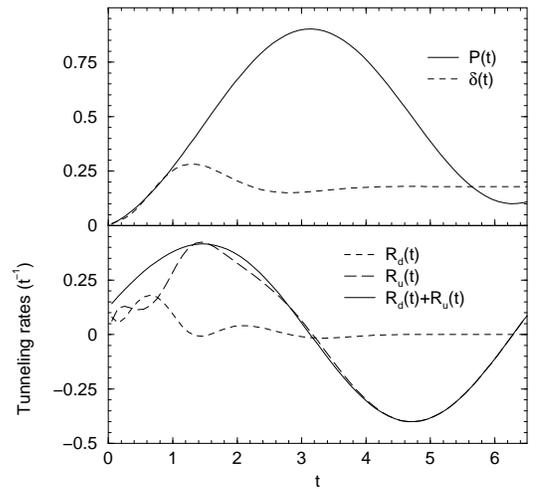}
\end{center}
\end{minipage}
\end{figure}

\end{document}